\documentclass[]{article}
\usepackage{graphicx} 
\usepackage{amscd}        
\usepackage{amssymb,amsmath,color}
\usepackage[a4paper, total={6.5in, 10in}]{geometry}
\numberwithin{equation}{section}
\usepackage{authblk}
\usepackage{graphicx, caption, subcaption}

\title{The Harmonic Oscillator Potential Perturbed by a Combination of Linear and Non-linear Dirac Delta Interactions with Application to 
Bose-Einstein Condensation}

\author[1]{Cenk Aky\"{u}z}
\author[2]{Fatih Erman}
\author[3]{Haydar Uncu}
\affil[1]{Department of Physics, Aydın Adnan Menderes University, Efeler, 09100, Aydın, Türkiye}
\affil[2]{Department of Mathematics, \.{I}zmir Institute of Technology, Urla, 35430, \.{I}zmir, Türkiye}
\affil[3]{Department of Mechatronics, Turkish German University, Beykoz, \.{I}stanbul, Türkiye}
\affil[1]{cenk.akyuz@adu.edu.tr}
\affil[2]{fatih.erman@gmail.com}
\affil[3]{haydar.uncu@tau.edu.tr}

\begin{document}

\maketitle

\begin{abstract}
In this paper, we study the bound state analysis of a one dimensional nonlinear version of the Schr\"{o}dinger equation for the harmonic oscillator potential perturbed by a $\delta$ potential, where the nonlinear term is taken to be proportional to $\delta(x) |\psi(x)|^2 \psi(x)$. The bound state wave functions are explicitly found and the bound state energy of the system is algebraically determined by the solution of an implicit equation. Then, we apply this model to the Bose-Einstein condensation of a Bose gas in a harmonic trap with a dimple potential. We propose that the many-body interactions of the Bose gas can be effectively described by the  nonlinear term in the Schr\"{o}dinger equation. Then, we investigate the critical temperature, the condensate fraction, and the density profile of this system numerically.    
\end{abstract}

\section{Introduction}

The Dirac delta potentials generally known as point/contact interactions are not a new subject and they are well-known exactly solvable models extensively studied from various points of view in the literature. They are considered to be ideal models for some physical systems in atomic, nuclear, and condensed matter physics, where the particle's de Broglie wavelength is much larger than the range of the potential (see \cite{Demkov} and the references in the monograph \cite{Albeverio}). For instance, the Kronig-Penney model \cite{KronigPenney} is one of such well-known ideal models, where the impurities in solids are described by periodic delta potentials to explain the energy band gap structures in metals.

The Hamiltonian operator associated with such interactions has been described in a more rigorous way in mathematics literature \cite{Albeverio, AlbeverioKurasov} and they have been also studied in two and three dimensions for a better understanding of the concept of renormalization, which was historically first introduced in quantum field theories, see e.g., \cite{Jackiw}. 

The applications of Dirac delta potentials are not restricted to the linear Schr\"{o}dinger equation. A non-linear version of the Schr\"{o}dinger equation 
\begin{eqnarray}
    -\frac{d^2 \psi}{dx^2}- \Omega |\psi(x)|^{a} \delta(x) \psi(x)=E \psi(x) \;, \label{nonlineardeltaschrodinger}
\end{eqnarray}
has been suggested to describe the optical wave propagation in a one-dimensional linear medium which contains a narrow Kerr-type non-linear strip within scalar approximation \cite{Moya, Molina}. Within this model, $\Omega$ is interpreted as the opacity and $a$ is the non-linear exponent. The same form of the above equation can be considered as a toy model in quantum many-body problems in which the many-body  interactions are replaced by an effective nonlinear system using mean field approximation. In this case, the choice of the non-linear exponent $a=2$ effectively describes an electron propagating in a one-dimensional linear medium that includes a strongly interacting vibrational impurity at the origin \cite{CMT93}. We will follow the same interpretation of the non-linear delta potentials as an effective model of many-body interactions in this paper. The bound state solutions to the above equation have been discussed in \cite{Molina} and it has been extended to the multi-center case in our earlier work \cite{UncuErman}. The time dependent version of this problem in mathematics literature has been investigated in the context of nonlinear Schr\"{o}dinger equation with concentrated nonlinearities in various dimensions and existence/uniqueness of the solutions and its blow-up analysis have been studied in detail in \cite{Cacciapuoti, Carlone, Cacciapuoti2}.

The attractive Dirac $\delta$ potentials have been also proposed to study the thermodynamic properties of a Bose gas and in particular Bose-Einstein condensation (BEC) \cite{JankeCheng, IoriattiRosaHipolito}. In  more recent works \cite{Huncu, Huncu2}, they have been proposed to describe the dimple potentials in a harmonically trapped Bose gas, and the potential of the entire system is given by  $\frac{1}{2}m\omega^2x^2- \frac{\hbar^2 \sigma}{2m} \delta(x-a)$. Here $\sigma>0$ is a coupling representing the strength of the dimple potential, $a$ represents the location of dimple potential and the coefficient of  $\delta$ function is chosen as, $- \frac{\hbar^2 \sigma}{2m}$ for the sake of calculational convenience. It is well-known that this model can be solved analytically \cite{Grosche,Fassari1, Fassari2, Ersan2, Fassari3, Fassari4, Janev}, that is, one can explicitly determine the eigenfunctions of the Hamiltonian associated with this potential and its eigenvalues can be found by solving an implicit equation numerically. From these eigenvalues, the critical temperature of the BEC can be found and all the other thermodynamic properties of the system have been analyzed in detail and it is shown that an attractive dimple potential at the center of the harmonic trap leads to an increase in critical temperature, condensate fraction at a given temperature below critical temperature and density of boson gas  around the center of the trap \cite{Huncu}. In a similar work same authors investigated the effect of an off dimple \cite{Huncu2}. The critical temperature and condensate fraction at a given temperature also increase in this case, however, the amount of increase is smaller. In the same study, it is also shown that the peak of density profile can be translated from the center to the position of the off-center dimple for strong dimples. It is important to emphasize that the many body interactions of the Bose gas are neglected in this model \cite{Huncu}.

Dimple type potentials are utilized in different areas of physics like quantum tweezers for atoms \cite{Diener}, atom lasers \cite{Stellmer}, and soliton-sound interaction \cite{Parker}. However, they have the most promising applications  in the field of BEC. They have been introduced for the purpose of increasing the phase space density  by modifying the shape of trapping potentials \cite{Kurn, Pinkse, Weber, Ma}. It has been shown in \cite{Kurn} that the phase-space density can be increased by an arbitrary factor by using a small dimple potential at the equilibrium position of the harmonic trap. This has been demonstrated by a cesium BEC in \cite{Weber}, where a tight dimple potential has been used. Such potentials have been also proposed for efficient loading and fast evaporative cooling to produce large condensate \cite{Comparat} and for measurement-enhanced determination of BEC phase transitions \cite{Bason}, and as a pulsed output coupler which coherently extracts atoms
from a condensate \cite{Mewes,Andrews,Bloch}. Moreover, they have been used to create continuous-wave condensate of strontium atoms that lasts indefinitely \cite{Chen}. The experimental developments have attracted theoretical studies and several researchers have constructed models for various phenomena about boson gas in a harmonic trap with an additional dimple potential. For instance, an interesting model has been developed in \cite{Dutta}, which successfully describes the kinetics of BEC in a dimple potential by considering Bose gas in a harmonic trap as a reservoir. The statistics and the dynamics of quasi one-dimensional Bose-Einstein condensate of neutral $^{87}$ Rb atoms in a harmonic and dimple trap have been studied in \cite{Akram}. The effect of a dimple potential on ground-state properties of a quasi- one-dimensional condensate with two-and three-body interactions has been also worked out in \cite{Karabulut}. Furthermore, the chaotic dynamics in a condensate  has been analyzed using dimple potential in \cite{Sakhel}.

In this paper, we study  a non-linear model by adding an effective  repulsive non-linear $\delta$ term to the harmonic oscillator potential given by
\begin{equation}\label{newmodel}
    V(x)=\frac{1}{2}m\omega^2x^2+f(\vert\psi(x)\vert)\delta(x) \;,
\end{equation}
where we choose the function $f$ as
\begin{equation}\label{newmodelfunc} 
f(|\psi(x)|)=  -\frac{\hbar^2 }{2m} (\sigma-\alpha \vert \psi(x)\vert ^2) \;,
\end{equation}
with $\sigma>0$ and $\alpha>0$. The Schr\"{o}dinger equation with such an effective potential can be considered as a non-linear eigenvalue problem. 
We believe that this system can be realized as a simple model for position dependent magnetic Feshbach resonances. It is well-known that the coefficient of the non-linear term $|\psi|^2 \psi$ in the Gross–Pitaevskii equation is proportional to the $s$-wave boson-boson scattering length:
\begin{equation*}
    \left(-\frac{\hbar^2}{2m} \nabla^2 +V(\mathbf{r}) + \frac{4\pi \hbar^2 a_s}{m} |\psi(\mathbf{r})|^2 \right) \psi(\mathbf{r}) = \mu \psi(\mathbf{r}) \;,
\end{equation*}
where $\mu$ is the chemical potential and the external potential $V$ in our case involves a harmonic trapping potential perturbed by a linear delta potential in modeling the dimple potential. In ultra-cold atomic experiments, the resonance is controlled by a magnetic field. Then, the scattering length $a_s$ additional to the background scattering length is of the form $\frac{\Delta}{B - B_0}$, where $\Delta$ is the resonance width. If magnetic field is taken as a position dependent function and then the expression for the scattering length near resonance can be approximated by a delta function. Our model here might be a possible realization of the one-dimensional version of this. In this work, we obtain the solution for the bound states and derive an implicit equation for the bound state energies associated with the potential given in (\ref{newmodel}) for a general form of $f$ and then consider the special choice (\ref{newmodelfunc}) for $f$. The results are interesting in the sense that for particular values of $\sigma$ and $\alpha$ there emerge extra bound state energies. We use the obtained results to give an effective model for one-dimensional Bose gas in a harmonic trap with a dimple potential as a possible physical application.

After analyzing the bound state spectrum of this model, we numerically determine the critical temperature, density profiles, and condensate fractions for different values of $\sigma$ and $\alpha$, representing the strength of the dimple and interactions between the particles, respectively. All the numerical computations in this paper have been performed by Mathematica.

The paper is organized as follows: In section \ref{genharmonic}, we present the solution of the ``non-linear" eigenvalue problem for the potential given in equations \eqref{newmodel} and \eqref{newmodelfunc}. In Section \ref{apptobose} we apply the results of the model to Bose gas in a harmonic trap with a dimple in one dimension and use the obtained bound state energies and wave functions to estimate the thermodynamic properties of  Bose gas at equilibrium. We finally give the conclusions.

\section{The Harmonic Potential Perturbed by a Combination of Linear and Non-linear Dirac Delta Potential \label{genharmonic} }

The nonlinear version of the time independent Schr\"{o}dinger equation for the problem that we consider here is given by
\begin{equation}\label{TISDwD}
    -\frac{\hbar^2}{2m}\frac{d^2\psi{(x)}}{dx^2}+\left[\frac{1}{2}m\omega^2x^2+f(|\psi(x)|)\delta(x)\right]\psi{(x)}=E\psi{(x)} \;,
\end{equation}
where $f$ is given in (\ref{newmodelfunc}). At first glance, one may think that this equation is equivalent to the linear problem since the coefficient of $\delta$ is a constant thanks to the shifting property of delta function, which may then reduce the problem to the standard one. However, this argument is not correct because the constant $-\frac{\hbar^2 }{2m} (\sigma-\alpha \vert \psi(0)\vert ^2)$ involves the unknown solution itself.

In order to find the particular bound state solution to this problem, we first consider the above non-linear equation for $x\neq0$:
\begin{equation}\label{TISD}
    -\frac{\hbar^2}{2m}\frac{d^2\psi{(x)}}{dx^2}+\frac{1}{2}m\omega^2x^2\psi{(x)}=E\psi{(x)} \;.
\end{equation}
Defining dimensionless variables $z:=\sqrt{\frac{2m\omega}{\hbar}}x$,
$\xi:=\frac{E}{\hbar\omega} -\frac{1}{2}$, equation \eqref{TISD} can be expressed simply as
\begin{equation}\label{reducedTISD}
    \frac{d^2\psi{(z)}}{dz^2}+\left(\xi+\frac{1}{2}-\frac{1}{4}z^2\right)\psi{(z)}=0 \;.
\end{equation}
The linearly independent solutions of the above equation are well-known and given by the parabolic cylinder functions $D_{\xi}(z)$ and $D_{\xi}(-z)$ \cite{lebedev}. One can express the parabolic cylinder functions in terms of the Hermite functions $H_{\xi}(z)$ \cite{lebedev}:
\begin{equation}\label{parabolic cylinder function}
    D_{\xi}(z)=2^{-\xi/2}\exp{(-z^2/4)}H_{\xi}(z/\sqrt{2}) \;,
\end{equation}
where $H_{\xi}(z)$ is defined by the confluent hyperbolic functions
\begin{equation}\label{Hermite}
   H_{\xi}(z)=2^{\xi}\left[\frac{\sqrt{\pi}}{\Gamma{(\frac{1-\xi}{2})}}{_1F_1\left(-\frac{\xi}{2},\frac{1}{2};z^2\right)}
                          -\frac{\sqrt{2\pi}z}{\Gamma{(-\frac{\xi}{2})}}{_1F_1\left(\frac{1-\xi}{2},\frac{3}{2};z^2\right)}\right] \;.
\end{equation}
Here $\Gamma(z)$ is the well-known gamma function. 
One can easily show that $D_{\xi}(z)$ and $D_{\xi}(-z)$ are two linearly independent solutions to equation (\ref{reducedTISD}) by simply calculating their Wronskian $W_D$ at $z=0$ without loss of generality. Using the values of the parabolic cylinder function at the origin \cite{lebedev},
\begin{equation}\label{parat0}
    D_{\xi}(0)=\frac{2^{\xi/2}\sqrt{\pi}}{\Gamma{(\frac{1-\xi}{2})}} \;,
\end{equation}
and
\begin{equation}\label{deroarat0}
    \frac{dD_{\xi}(z)}{dz} \Biggr\rvert_{z=0}=-\frac{\sqrt{2\pi} \, 2^{\xi/2}}{\Gamma(-\frac{\xi}{2})} \;,
\end{equation}
we find
\begin{equation}\label{wronskianz0}
      W_{D}:= W(D_{\xi}(z), D_{\xi}(-z);0) = -2D_{\xi}(0)\frac{dD_{\xi}(z)}{dz} \Biggr\rvert_{z=0} = \frac{2^{\xi+3/2}\pi}{\Gamma{(\frac{1-\xi}{2})}\Gamma{(-\frac{\xi}{2})}} \;.
\end{equation}
%
%
The asymptotic behavior of the parabolic cylinder functions $\lim_{z\rightarrow{\infty}}{D_{\xi}(-z)} = \infty$ and $\lim_{z\rightarrow{-\infty}}{D_{\xi}(z)} = \infty$ forces us to choose the bound state solutions of the equation (\ref{reducedTISD}) in the following way:
\begin{equation}\label{eigenfunctionswocont}
    \psi{(z)}=\left\{%
\begin{array}{ll}
    a \, D_{\xi}(-z) & \hbox{if $z<0$} \\
    b \, D_{\xi}(z) & \hbox{if $z>0$.} \\
\end{array}%
\right.
\end{equation} 
The continuity of solution at $z=0$ yields $a=b$. Using the formula 7.711.3 in \cite{GradRyznik} for the integral of the absolute square of parabolic cylinder function
\begin{equation}
    \int_{0}^{\infty} |D_{\xi}(x)|^2 d x = \pi^{1/2} 2^{-3/2} \left(\frac{\psi^{(0)}(\frac{1}{2}-\frac{\xi}{2}) -\psi^{(0)}(-\xi/2)}{\Gamma(-\xi)} \right) \;,
\end{equation}
where $\psi^{(0)}$ is the digamma function, $a$ can be found from the normalization condition, given by   
\begin{equation}\label{normalization}
a=\left[ \sqrt{\frac{\pi}{2}} \frac{\psi^{(0)}(\frac{1-\xi}{2}) -\psi^{(0)}(-\frac{\xi}{2})}{\Gamma(-\xi)}\right]^{-1/2} \;.
\end{equation} 
The derivative of the wave function is discontinuous at the origin due to the delta function. The amount of this discontinuity can be found by integrating the Schr\"{o}dinger equation around a small $\epsilon$ neighborhood of the origin
and taking the limit as $\epsilon \to 0^+$
\begin{equation}\label{jump}
    \lim_{\epsilon{\rightarrow{0^+}}}\left\{\int_{-\epsilon}^{+\epsilon}{\frac{d^2\psi{(z)}}{dz^2}dz}
    +\int_{-\epsilon}^{+\epsilon}{\left(\xi+\frac{1}{2}-\frac{z^2}{4}\right)}\psi{(z)}dz
    -\int_{-\epsilon}^{+\epsilon}{g(|\psi{(z)}|)\delta{(z)}\psi{(z)}dz}\right\}=0 \;,
\end{equation}
where
\begin{equation}\label{ftog}
g(|\psi{(z)}|) = \sqrt{\frac{2m}{\hbar^3\omega}}{f}(|\psi{(z)}|) = -\sqrt{\frac{\hbar}{2 m \omega}}\left(\sigma- \alpha |\psi(z)|^2 \right) \,.
\end{equation}
Here $ \psi(x(z)) \equiv \psi{(z)}  $ for the sake of notational simplicity. After some algebra and using the solution (\ref{eigenfunctionswocont}), the jump discontinuity condition \eqref{jump} gives
\begin{equation}\label{eigenvalue}
    -g(|\psi{(0)}|)D_{\xi}^2(0)=W_{D} \;.
\end{equation}
If $D_{\xi}(0) \neq 0$, the explicit form of the above equation in terms of $\xi$ is given by
\begin{equation}  \label{eigenergy}
    \frac{2^{3/2} \Gamma(\frac{1-\xi}{2})}{\Gamma(-\xi/2)} = \sqrt{\frac{\hbar}{2m \omega}} \Bigg[\sigma-\frac{\alpha \sqrt{\pi} 2^{\xi+\frac{1}{2}}}{\Gamma^2(\frac{1-\xi}{2})} \bigg(\frac{\Gamma(-\xi)}{\psi^{(0)}(\frac{1-\xi}{2})-\psi^{(0)}(-\xi/2)}\bigg) \Bigg] \;.
\end{equation}
The roots $\xi$ of this equation partly determine the bound state energies of our system through the relation $E=(\xi+\frac{1}{2}) \hbar \omega$. The wave function associated with these bound state energies is given by the parabolic cylinder functions (\ref{eigenfunctionswocont}). On the other hand, if $D_{\xi}(0)=0$, then $(1-\xi)/2$ must be poles of the gamma function according to equation (\ref{parat0}), that is, $\xi=2n+1$ for $n=0,1,2,\ldots$. This gives us further bound state solutions of our non-linear problem, given by $E=(2n+1+\frac{1}{2})\hbar \omega$, which exactly coincides with the odd eigenvalues of the harmonic oscillator potential. In this case, the solution is given by $2^{-n+\frac{1}{2}} \exp(-z^2/4) H_{2n+1}(z/\sqrt{2})$ for $n=0,1,2,\ldots$.

In the case $D_{\xi}(0)\neq 0$, we need to solve the equation (\ref{eigenergy}). For the given values of $\alpha$ and $\sigma$, the roots $\xi$ of this equation are rather hard to find analytically. Nevertheless, we may solve this equation numerically for particular values of $\alpha$ and $\sigma$ and the parameters $m$ and $\omega$. The case $\alpha=0$ in equation (\ref{eigenergy}) reduces to the same equation for the eigenvalues of the harmonic oscillator potential decorated with a linear $\delta$ potential, which is well known in the literature \cite{Huncu, Grosche, Fassari1}. However, the effect of the nonlinear term on the energy eigenvalues of harmonic oscillator perturbed by Dirac delta potential has not been studied so far to the best of our knowledge.

\begin{figure}[h!] \begin{center}
\includegraphics[scale=0.60]{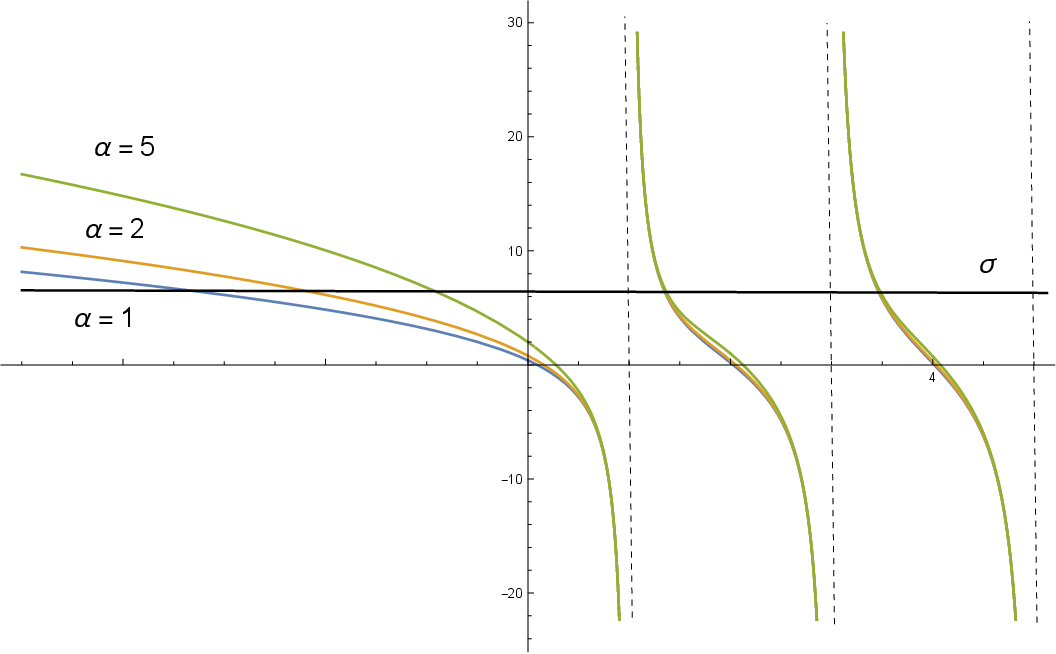}
\caption{The graph of the function $\frac{4\Gamma(\frac{1-\xi}{2})}{\Gamma(-\xi/2)} + \frac{\alpha \sqrt{2\pi} 2^{\xi}}{\Gamma^2(\frac{1-\xi}{2})} \bigg(\frac{\Gamma(-\xi)}{\psi^{(0)}(\frac{1-\xi}{2})-\psi^{(0)}(-\xi/2)}\bigg) $ with respect to $\xi$ for $\alpha=1,2,5$ units. Here $\hbar=m=\omega=1$ units for simplicity. The fixed value of $\sigma$ is shown as a horizontal line.}
\label{bss1}
\end{center}
\end{figure}

\begin{figure}[h!] \begin{center}
\includegraphics[scale=0.50]{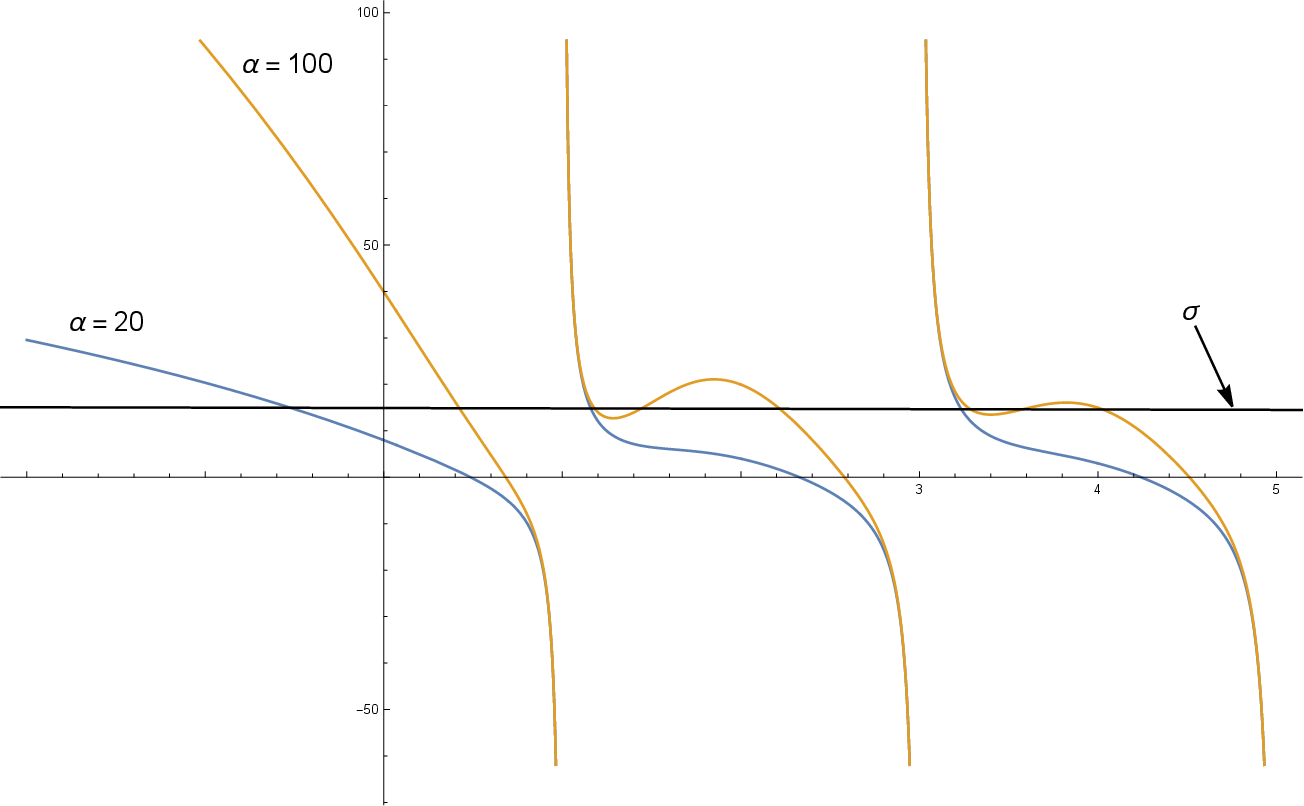}
\caption{The graph of the function $\frac{4\Gamma(\frac{1-\xi}{2})}{\Gamma(-\xi/2)} + \frac{\alpha \sqrt{2\pi} 2^{\xi}}{\Gamma^2(\frac{1-\xi}{2})} \bigg(\frac{\Gamma(-\xi)}{\psi^{(0)}(\frac{1-\xi}{2})-\psi^{(0)}(-\xi/2)}\bigg) $ with respect to $\xi$ for $\alpha=20, 100$ units. Here $\hbar=m=\omega=1$ units for simplicity. The fixed value of $\sigma$ is shown as a horizontal line.}
\label{bss12}
\end{center}
\end{figure}

We first consider the equation (\ref{eigenergy}) graphically to show how the bound state energies of the system change as a function of the couplings $\alpha$ and $\sigma$ before discussing the solutions numerically. It is useful to write equation (\ref{eigenergy}) by moving the second term on the right hand side to the left hand side and keeping the $\sigma$ term fixed on the right hand side. Then, if we plot the left hand side of this new equation as a function of $\xi$ for given values of $\alpha=1,2,5$ units with $\hbar=m=\omega=1$, the intersections of this graph with fixed values of $\sigma$ helps us to analyze how the bound state energies change with $\sigma$ (see Figure \ref{bss1}). Due to the poles of the gamma functions at negative integers, the vertical asymptotes in this Figure correspond to the odd values of $\xi$, that is, to the bound state energies $E= (2n+1+\frac{1}{2})\hbar \omega$ with $n=0,1,2,\ldots$. It is easy to see from Figure \ref{bss1} that the rest of the bound state energies are interlaced between these vertical asymptotes and as we increase the values of $\sigma$, the ground state energy decreases, whereas the interlaced excited bound state energies decrease till to the closest vertical asymptote. If $\alpha$ is sufficiently large, two more excited bound states will emerge between these vertical asymptotes, as can be seen from Figure \ref{bss12}. It is also important to realize that the number of excited states goes to one as we go to the higher excited states.   

\begin{figure}[h!] \begin{center}
\includegraphics[scale=0.60]{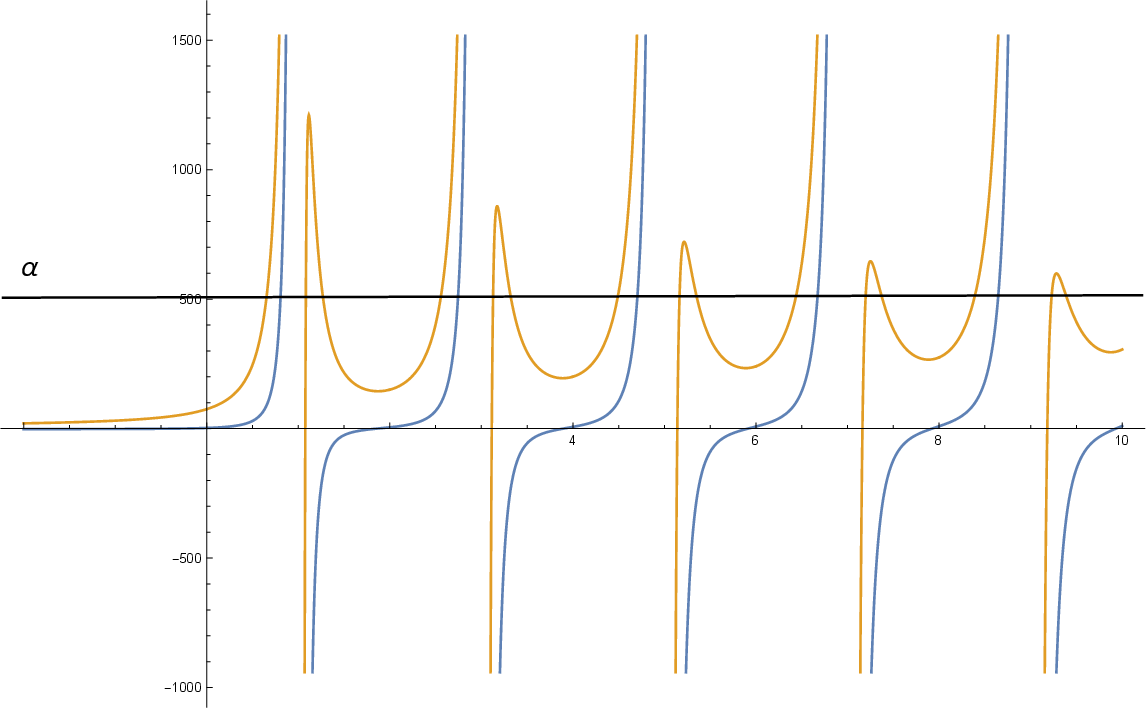}
\caption{The graph of the function $\frac{\Gamma^2(\frac{1-\xi}{2})}{2^{\xi} \sqrt{2\pi}\Gamma(-\xi)} \left(\psi^{(0)}(\frac{1-\xi}{2})-\psi^{(0)}(-\xi/2)\right)\left(\sigma-\frac{4 \Gamma(\frac{1-\xi}{2})}{\Gamma(-\xi/2)}\right)$ with respect to $\xi$ for $\sigma=1$ units (blue curve) and $\sigma=30$ units (yellow curve). Here $\hbar=m=\omega=1$ units for simplicity.} \label{bss2}
\end{center}
\end{figure}

\begin{figure}[h!] 
\begin{center} 
\includegraphics[scale=0.6]{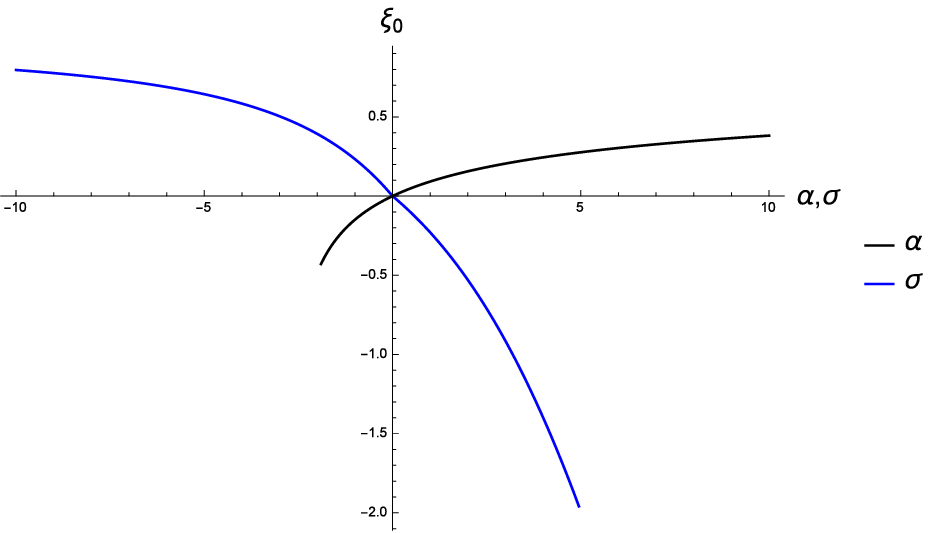}
\caption{The change of ground state energy eigenvalue with respect to $\alpha$ and $\sigma$.} \label{E0vsLambSig}
\end{center}
\end{figure}

\begin{figure}[h!] 
\begin{center} 
    \includegraphics[scale=0.6]{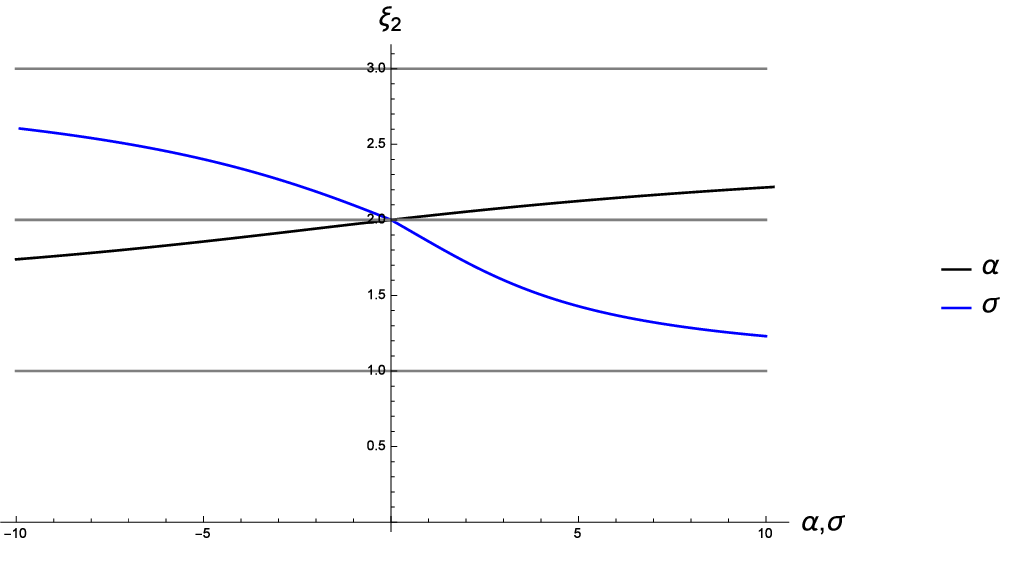}
\caption{The change of the second excited state energy eigenvalue with respect to $\alpha$ and $\sigma$.} \label{E2vsLambSig}
\end{center}
\end{figure}

If we keep $\alpha$ alone in equation (\ref{eigenergy}), and plot the other side of the equation as a function of $\xi$, the situation is similar to the one in Figure \ref{bss1} and Figure \ref{bss12}, that is, the excited bound state energies are interlaced  between the vertical asymptotes at the odd integer values of $\xi$. If $\sigma$ is sufficiently large, the number of bound states between the vertical asymptotes changes from one to three, as can be shown in Figure \ref{bss2}. From this Figure, we have seen that the peak between the vertical asymptotes decreases so that the number of excited bound states goes to one as we go to the higher excited states for a fixed value of $\alpha$.  
In order to illustrate how the bound state energies change with respect to $\alpha$ and $\sigma$, we plot the numerically calculated ground state and second excited state energies with respect to $\alpha$ and $\sigma$ in Figure \ref{E0vsLambSig} and Figure \ref{E2vsLambSig}. The vertical axes in these Figures are $\xi_0:=\frac{E_0}{\hbar\omega} -\frac{1}{2}$ and $\xi_2:=\frac{E_2}{\hbar\omega} -\frac{1}{2}$, respectively. When we vary $\alpha$ we take $\sigma=0$ and when we vary $\sigma$ we take $\alpha=0$.
First, we present the variation of $\xi_0$ with respect to $\alpha$ and $\sigma$ . The behaviour of the ground state energy with respect to $\alpha$, for $\alpha<0$ is interesting, because it rapidly decreases as $\alpha$ varies from zero to $-2$ and then it disappears. This is the reflection of the fact that the equation \eqref{eigenergy} has no solution $\xi$ for $\sigma=0, \alpha  \leq -2$.

Similarly, the behaviour $\xi_2$ with respect to $\alpha$ and $\sigma$ is depicted in Figure \ref{E2vsLambSig}. The rate of change of $\xi_2$ with respect to $\sigma$ is considerably larger than that of $\xi_2$ with respect to $\alpha$. In both cases, $\xi_2$ approximates to 1 as  positive $\sigma$ values increase and  negative $\alpha$  values decrease. Similarly, it approximates to 3 for the symmetric case. The horizontal asymptotes $1$ and $3$ correspond to the odd excited eigenstates of the harmonic oscillator potential for $n=1$ and $n=3$, respectively. The eigenvalues corresponding to the odd eigenstates of the harmonic oscillator do not change with  either $\sigma$  or $\alpha$ as mentioned above. 

We will use these numerically obtained bound state energies of our nonlinear problem to calculate some thermodynamic properties of boson gas, such as the critical temperature, and the condensate fraction of a boon gas in the presence of the effective potential in the next section. 

\section{Application to Bose Gas \label{apptobose}} 

In this section, we investigate how the critical temperature, the condensate fraction and the density profile of Bose gas in a harmonic trap change with respect to the strength of the linear delta potential for particular values of the coupling constant $\alpha$ of the non-linear delta potential. As mentioned in the previous section, we model dimple potential by the term $-\frac{\hbar^2}{2m} \sigma \delta(x)$ in the Schr\"{o}dinger equation. The experimental studies in three dimensions with a dimple potential \cite{Stellmer, Kurn} show that the density of boson gas around the center of the harmonic trap increases considerably in the presence of an attractive dimple potential centered at the origin. Therefore, it is not realistic to ignore interactions for a boson gas in a harmonic trap with a dimple potential. To take interactions into account, we propose a simple effective model by adding a concentrated non-linear term to the Schr\"{o}dinger equation.

It is convenient to define a dimensionless variable to represent the strength of the dimple potential
\begin{equation}\label{Lambda}
\Lambda= \sigma\sqrt{\frac{\hbar}{m \omega} } \;.
\end{equation}
We shall use the following experimental values  $m=23$ amu $(^{23}Na)$ for the mass of bosons, $ \omega=2 \, \pi \, \times 21$ Hz \cite{hau} for the frequency of harmonic trap in the below numerical calculations. Moreover, using the experimental values in the references \cite{Ma} and \cite{Garret}, we give an estimate for the range of the possible values of $\Lambda$. The maximum depth of the dimple potential is given as $4 \mu K \, \textrm{k}_b $, where $\textrm{k}_b$ denotes the Boltzmann constant, and the width of the dimple varies from 1 $\mu m$ to $160$ $\mu m$ in \cite{Ma}). The authors in \cite{Garret} have created two dimples: one with a maximum depth of $210 \, \textrm{n} K\, \textrm{k}_b$  and a width of  $ 11 \mu m$ and  another one  $1610 \, \textrm{n} K \, \textrm{k}_b$ and a width of  $ 32 \mu m$. From these experimental values, the range of $\Lambda$ is approximately estimated between $0$ and $5 \times 10^2$.

In the rest of this section, we investigate numerically  the behavior of the thermodynamic parameters like the critical temperature, the chemical potential, and the condensate fraction by choosing $\alpha=10^6$, which is sufficient to see the effect of the non-linearity. 

\subsection{The Critical Temperature}
We first consider the change of critical temperature $T_c$ with respect to $\Lambda$ for different average numbers of particles for $\alpha=0$ and $\alpha=10^6$. 
The condition for the occurrence of BEC in the grand canonical ensemble is that the average number of particles in the excited states remains finite as the chemical potential $\mu$ approaches the ground state energy $E_0$ from below, that is, 
\begin{eqnarray} \label{criticalTcondition}
\lim_{\mu \to E_0} N_{ex} = \lim_{\mu \to E_0} \sum_{i=1}^{\infty} \frac{1}{e^{\beta(E_i-\mu)}-1} < \infty \;,
\end{eqnarray}
where $\beta= \frac{1}{\textrm{k}_b T}$ and $E_i$ is the energy of the state $i$. According to the Mermin-Wagner-Hohenberg theorem, one might believe that there will be no condensation at finite temperature in one dimension. However, the statement of this theorem is true under the assumption that the Bose gas is homogeneous, whereas our situation include the harmonic trapping potential which breaks this homogeneity \cite{Görlitz}. It is not obvious that BEC can still occur under an additional non-linear term to the Schr\"{o}dinger equation. However, as discussed in the previous section, the bound state energies asymptotically approach $E_n=(2n+1+\frac{1}{2})\hbar \omega$ for large values of $n$ and energies become doubly degenerate. Since this is exactly the bound state energies of the harmonic oscillator for odd eigenstates, it ensures the convergence of the above sum by simply the ratio test. Hence, the condition for the occurrence of BEC at finite temperature is satisfied. 

Then, the critical temperature $T_c$ can be obtained from the above equation by taking the limit $\mu \to E_0$ in equation (\ref{criticalTcondition}) and equating $N_{ex}$ to the total number of particles $N$:
\begin{eqnarray} \label{tcrit}
    N= \sum_{i=1}^{\infty} \frac{1}{e^{\beta_c(E_i-E_0)}-1} \;,
\end{eqnarray}
where $\beta_c=1/(k_b T_c)$. 
\begin{figure}[h!]
\begin{center}
    \includegraphics[scale=0.75]{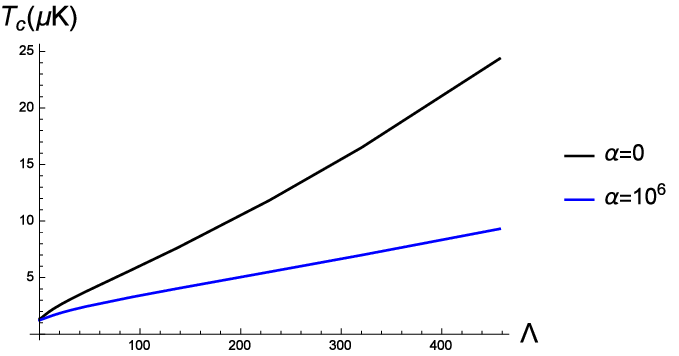}
\caption{The change of critical temperature $T_c$ with respect to $\Lambda$ for $\alpha=0$ (black curve) and $\alpha=10^6$ (blue curve).}
\label{TcvsLN4}
\end{center}
\end{figure}

Unfortunately, it is difficult to solve $T_c$ analytically from equation (\ref{tcrit}). For this reason, the critical temperature $T_c$ in terms of the dimensionless coupling constant $\Lambda$ of linear delta potential is found by solving equation (\ref{tcrit}) numerically by plugging the numerically obtained solutions of the energies $E_i=\hbar \omega (\xi_i + \frac{1}{2})$ from equation (\ref{eigenergy}) for a fixed value of the number of particles $N=10^4$ and two different values  $\alpha=0$ and $\alpha=10^6$. Then, the graphs of the critical temperature $T_c$ for $N=10^4$ with respect to $\Lambda$ are plotted as shown in Figure \ref{TcvsLN4}, for $\alpha=0$ and for $\alpha=10^6$. Here, the black curve corresponds to the change in $T_c$ for $\alpha=0$, where the non-linear interaction is ignored and the blue curve corresponds to the change in $T_c$ in the presence of non-linear interaction with the choice of coupling $\alpha=10^6$. What we have observed here is that 
$T_c$ increases almost linearly with increasing $\Lambda$ in both $\alpha=0$ and $\alpha=10^6$ cases. Such a behavior for $\alpha=0$ is actually expected and can be seen by the following argument. It is well-known that the energy gap of the harmonic oscillator potential perturbed by a linear $\delta$ potential between the excited and the ground state energies $E_i-E_0$ increases as the strength $\sigma$ of the linear $\delta$ potential increases (see the recent work \cite{Erman1} for a detailed discussion how the spectrum of some regular potentials changes under the perturbation of linear delta potential). Then, it follows from equation (\ref{tcrit}) that the critical temperature $T_c$ must increase with respect to $\Lambda$ as long as we keep the number of particles $N$ fixed. However, adding the non-linear interaction reduces the rate of change of the critical temperature by a considerable amount. The difference between the critical temperatures for $\alpha=0$ and $\alpha=10^6$ is getting smaller around $\Lambda=0$, see Figure \ref{TcvsLN4}.

\begin{figure}[h]
    \begin{subfigure}[b]{0.5\textwidth}
      \includegraphics[width=\textwidth]{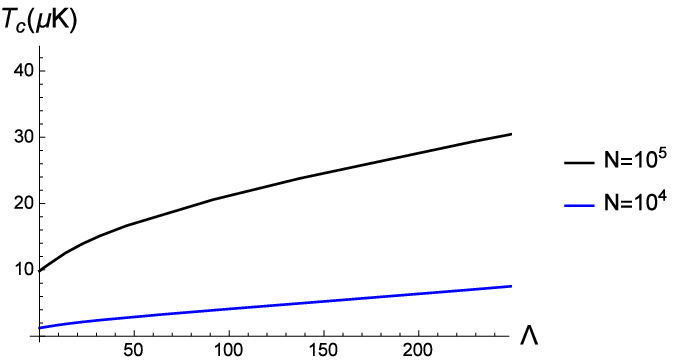}
      \caption{}
    \end{subfigure}
    \begin{subfigure}[b]{0.5\textwidth}
      \includegraphics[width=\textwidth]{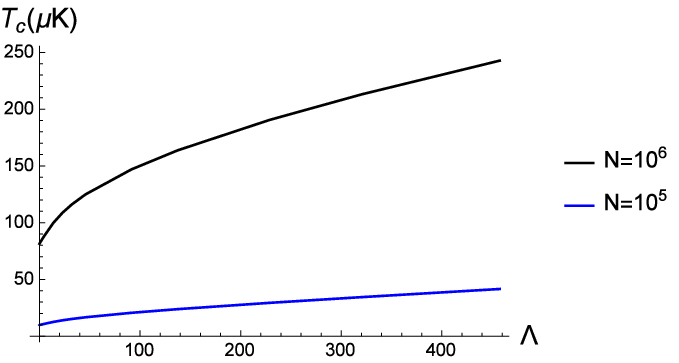}
      \caption{}
    \end{subfigure}
    \begin{subfigure}[b]{0.5\textwidth}
      \includegraphics[width=\textwidth]{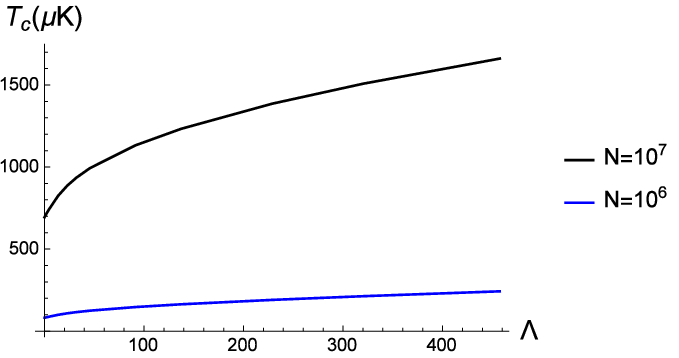}
      \caption{}
    \end{subfigure}
    \begin{subfigure}[b]{0.5\textwidth}
      \includegraphics[width=\textwidth]{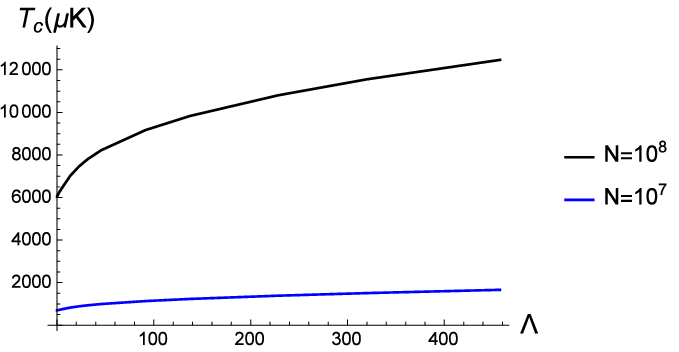}
      \caption{}
    \end{subfigure}    
    \caption{The critical temperature $T_c$ versus $\Lambda$ with $\alpha=10^6$ for different number of particles: (a) for $N=10^4$ and $N=10^5$ (b) for $N=10^5$ and $N=10^6$, (c) for $N=10^6$ and $N=10^7$, (d) for $N=10^7$ and $N=10^8$.}
\label{TcvsN4N6N8}
  \end{figure}

In order to see the effect of the number of particles $N$ on how the critical temperature changes with respect to $\Lambda$, we can perform similar calculations and the results are shown in Figure \ref{TcvsN4N6N8} for different values of $N$ with a fixed value of $\alpha$. As one can see from the graphs in Figure \ref{TcvsN4N6N8}, the critical temperature is sharply increased for a sufficiently large number of particles. As is well-known, the critical temperature of a non-interacting boson gas in a bare harmonic trap in one dimension is directly proportional to the particle number (see e.g., \cite{pethick}). This dependence seems to be preserved approximately in the presence of dimple and non-linear interactions.

\subsection{The Condensate Fraction}

\begin{figure}[h!]
    \begin{subfigure}[b]{0.5\textwidth}
      \includegraphics[width=\textwidth]{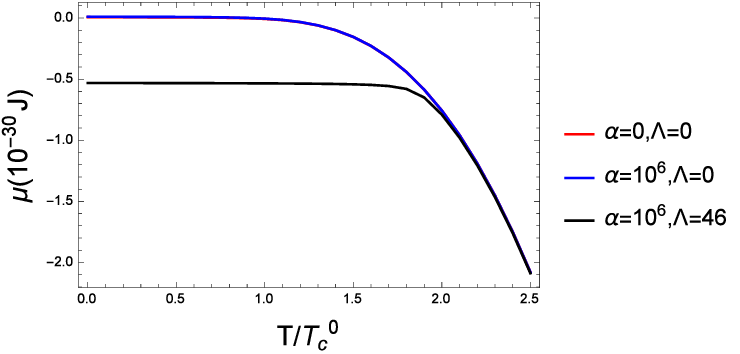}
      \caption{}
    \end{subfigure}
    \begin{subfigure}[b]{0.55\textwidth}
      \includegraphics[width=\textwidth]{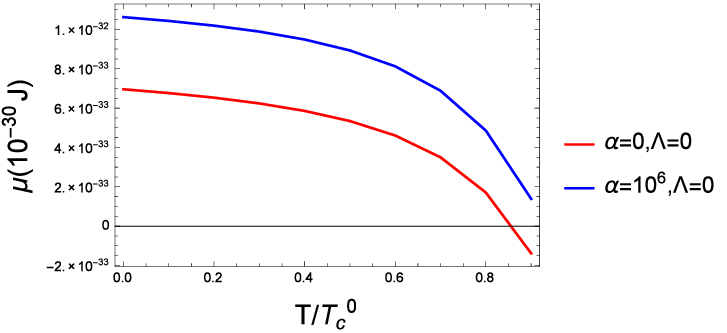}
      \caption{}
    \end{subfigure} 
    \caption{The chemical potential $\mu$ versus $T/T_c^0$ for different values of $\alpha$ and $\Lambda$.}
\label{muvsT}
  \end{figure}

We present in this section the variation of condensate fraction with temperature $T$ for different values of $\Lambda$ and $\alpha$. To do this, we first plot the behaviour of the chemical potential $\mu$ with respect to temperature for $N=10^4$ and for different $\alpha$ and $\Lambda$ values. The chemical potential at a given temperature and for a fixed number of total particles is calculated by numerically solving the following equation for $\mu$
\begin{equation}\label{muvst}
N = \sum_{i=0}^{\infty} {{1}\over{e^{\beta (E_i-\mu)}  -1}} \;,
\end{equation}
where the sum goes over all states including the ground state. The numerical solution $\mu$ for different values of temperature $T$ is then obtained and the behaviour of $\mu$ with respect to temperature is plotted by curve fitting given in Figure \ref{muvsT}. We scale the temperature axis with the critical temperature  $T_c^0$ of a boson gas in a harmonic trap without a dimple potential.  Here we fix the number of particles and take $N=10^4$. Figure \ref{muvsT} (a) shows the variation of $\mu$ with the change of temperature for three different pairs of couplings: $\alpha=0$, $\Lambda=0$; $\alpha=10^6$, $\Lambda=0$ and $\alpha=10^6$, $\Lambda=46$, where the temperature goes up to $2.5\, T_c^0$. In this figure, the chemical potentials for $\alpha=0$, $\Lambda=0$ and $\alpha=10^6$, $\Lambda=0$ can not be distinguished from each other for all temperature values. However, this is misleading because $\mu$ decreases rapidly with temperature and the curves for different $\alpha$ and $\Lambda$ values approximate each other for large temperature values. If we zoom the range of the graph around the origin, we obtain the graphs in Figure \ref{muvsT}(b),  which shows the splitting of the curves for two pairs of choices of the couplings.

\begin{figure}[h!]
\begin{center}
    \includegraphics[scale=0.75]{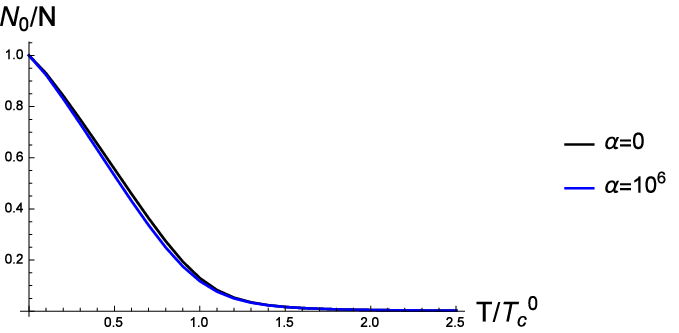}
\caption{The change of condensate fraction $N_0/N$ with respect to temperature in a harmonic trap without a dimple potential $\Lambda=0$. The temperature axis is scaled by $T_c^0$ for $N=10^4$. The black curve shows the change for $\alpha=0$, the blue curve shows the change for $\alpha=10^6$.}
\label{cfL0N4alfcomp}
\end{center}
\end{figure}

\begin{figure}[h!]
\begin{center}
\includegraphics[scale=0.75]{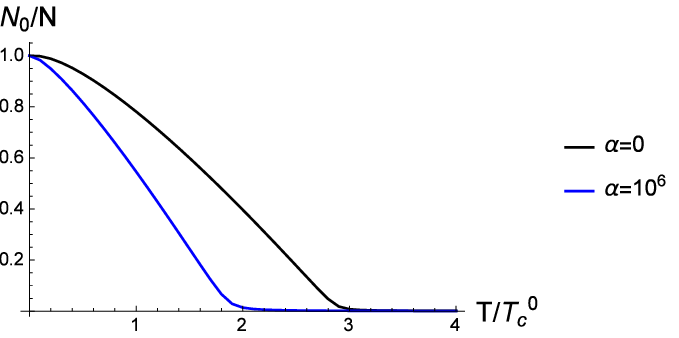}
\caption{The change of condensate fraction $N_0/N$ with respect to temperature in a harmonic trap with a dimple potential ($\Lambda=46$) for $N=10^4$. The temperature axis is scaled by $T_c^0$. The black curve shows the change for $\alpha=0$, the blue curve shows the change for $\alpha=10^6$.}
\label{cfL6S7N4TcBHvsS0}
\end{center}
\end{figure}

In the thermodynamic limit $N \rightarrow \infty$ and $V \rightarrow \infty$ with the particle density held fixed, semi-classical approximation is applicable \cite{pethick} and the chemical potential becomes equal to the ground state energy as $T \to T_c$ from above and it remains fixed at this value below the critical temperature. However, for systems with a finite number of particles $\mu$ is always smaller than the ground state energy and its value can be found by solving Eq. \eqref{muvst} for each value of temperature numerically. Inserting this value at the given temperature to 
\begin{equation}
 N_0=\frac{1}{e^{\beta(E_0 -\mu)}-1}, 
\end{equation}
one can get the mean particle number in the ground state and then calculate condensate fraction $N_0/N$. We calculate $N_0/N$ through this method  for different values of $\Lambda$ and for $\alpha=0$ or $\alpha=10^6$ and present the results in Figure \ref{cfL0N4alfcomp} and Figure \ref{cfL6S7N4TcBHvsS0}.

The change of condensate fraction in a  harmonic trap without a dimple potential ($\Lambda=0$) is depicted in Figure \ref{cfL0N4alfcomp}, where we consider the cases $\alpha=0$ and $\alpha=10^6$. It turns out that only the nonlinear term which effectively describes the interaction between bosons is not sufficient to change the condensate fraction considerably.  

In Figure \ref{cfL6S7N4TcBHvsS0}, we repeat the similar computation by including the dimple potential with  $\Lambda=46$ for a boson gas in a harmonic trap and see that the presence of nonlinear interaction around the origin decreases the condensate fraction considerably in this case.

These results may be interpreted as: The particles of a dilute boson gas in a harmonic trap without a dimple potential do not condense around the origin in the sense that only the nonlinear interaction is not sufficient to change considerably the thermodynamic quantities like chemical potential and condensate fraction.
In other words the interactions between the bosons remain weak also around the origin and do not affect the values of thermodynamic quantities remarkably.
Whereas the boson gas in a harmonic trap with a dimple potential condenses spatially around the origin so much strong that, the interactions between particles can not be ignored and they cause considerable changes in thermodynamic quantities.

\begin{figure}[h!]
\begin{center}
\includegraphics[scale=0.75]{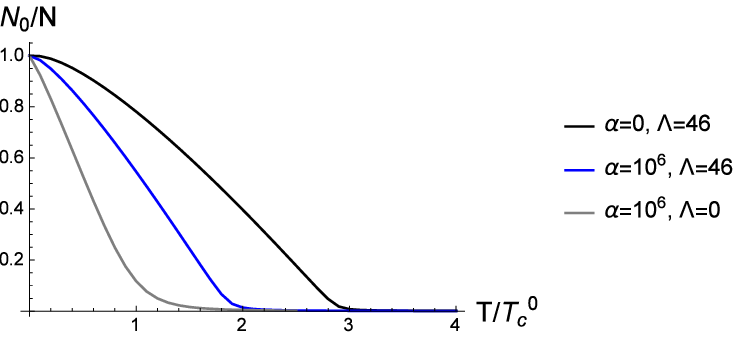}
\caption{The change of condensate fraction $N_0/N$ with respect to temperature in a harmonic trap.  The black and blue curves show the change for $\Lambda=46$ and $\alpha=0$, $\alpha=10^6$, respectively. The gray curve shows the change for $\alpha=10^6$ and $\Lambda=0$. }
\label{cfN4comp}
\end{center}
\end{figure}

It turns out that the presence of the dimple potential causes a substantial increase in the condensate fraction. On the other hand, the nonlinear repulsive interaction tends to decrease the condensate fraction, even if only the interaction around the origin is considered, as shown in Figure \ref{cfN4comp}. In other words, the nonlinear repulsive interaction between bosons and the dimple potential may be considered as two competitive effects. According to Figure \ref{cfN4comp}, the dimple potential causes an increase in condensate fraction, also in the existence of the nonlinear interaction, as one can see by comparing the blue curve ($\Lambda=46$, $\alpha=10^6$) with the gray one.

\begin{figure}[h!]
\begin{center}
    \includegraphics[scale=0.75]{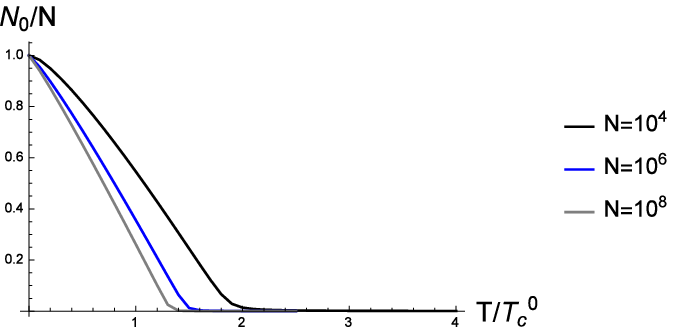}
\caption{The change of condensate fraction $N_0/N$ with respect to temperature for different total numbers of particles. The temperature axis is scaled by $T_c^0$. The black, blue and gray curves show the changes for $N=10^4$, $N=10^6$ and  $N=10^8$, respectively.}
\label{cfs7L6Ncomp}
\end{center}
\end{figure}

In the following figure, we present the variation of the condensate fraction of boson gases for $\Lambda=4.6$, $\alpha=10^6$ and with different total particle numbers. All curves are scaled by the critical temperature values $T_c^0$  corresponding to their particle numbers and then the curves are brought together in Figure \ref{cfs7L6Ncomp}. The critical temperatures are $T_c^0=1.30 \, \mu \textrm{K}$, $T_c^0=84.6 \, \mu \textrm{K}$ and $6.22 \, \textrm{mK}$, for $N=10^4$, $N=10^6$  and $N=10^8$, respectively. We see that the interactions give rises to larger decrease in the condensate fraction when the total particle number increases. This is expected, because as the particle number increases the gas around the origin becomes denser and interactions become dominant. The effective model we are using predicts this effect correctly, and gives an opportunity to estimate the condensate fraction also quantitatively as well.

\begin{figure}[h!]
\begin{center}
\includegraphics[scale=0.75]{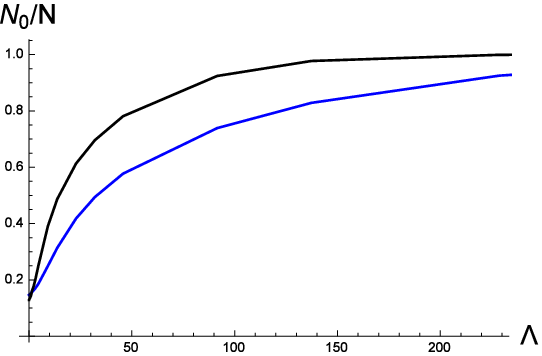}
\caption{The change of condensate fraction $N_0/N$ with respect to $\Lambda$. The black curve shows the change for $\alpha=0$, the blue curve shows the change for $\alpha=10^6$.}
\label{CFvsLamb}
\end{center}
\end{figure}

Finally, we discuss the change of condensate fraction at a constant temperature ($T=T_c^0$) with respect to $\Lambda$ and the strength of the dimple potential $\Lambda$. We observe  from Figure \ref{CFvsLamb} that as the strength of the
dimple potential increases, the condensate fraction rises remarkably. The nonlinear interaction makes the rate of change smaller but it does not eliminate the general behaviour. This may be an explanation for the growth of a Bose-Einstein condensate in a dimple trap without cooling. \cite{Garret}   

\subsection{The Density Profile}
The spatial distribution of particles in a gas is represented by a density profile. The density profile of the condensate phase of a Bose gas is given by the absolute square of the ground state wave function $\psi_0(x)$ of the single particle in some trapping potential
\begin{equation}
n_0(x)=\vert \psi_0 (x) \vert^2 \;.
\label{BECDP}
\end{equation}
Here all particles in the condensate are in the ground state of trapping potential. In this subsection, we numerically study the density profiles of the condensates by taking into account various interaction terms in our system.

One of the main advantages of the model under consideration is its exact solvability in the sense that the form of the bound state wave functions can be explicitly found and the energies can numerically be determined from an implicit equation. It is then possible to determine not only the density profile of the Bose-Einstein condensate but also of the boson gas. The  density profile of boson gas in one dimension is given by \cite{pethick}
\begin{equation}
n(x)= \sum_i f_{i} \vert \psi_{i} (x) \vert^2,
\label{bosonDP}
\end{equation}
where $\psi_{i}$ denotes the bound state wave functions corresponding to the energies $E_{i}$ and $f_{i}$ denotes the average number of of bosons in the state $i$, given by Bose-Einstein distribution:
\begin{equation}
f_{i}= {{1}\over{e^{\beta (E_{i}-\mu)}  -1} }.
\label{BEdist}
\end{equation}
The bound state wave functions of the model are formally given by equation \eqref{eigenfunctionswocont} in terms of parabolic cylinder functions \eqref{parabolic cylinder function} and the one corresponding to the odd eigenstates of harmonic oscillator potential. The normalization constant of the parabolic cylinder function) is given by (\ref{normalization}) in terms of parameters $\xi$ determined by the numerical solution of equation (\ref{eigenergy}). 

\begin{figure}[h!]
\begin{center}
\includegraphics[scale=0.75]{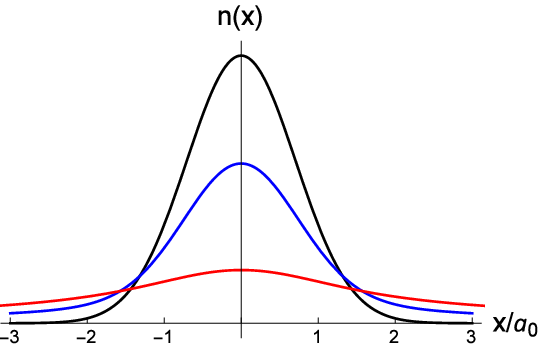}
\caption{The density profiles of boson gas at $T=0,\, T=(1/2)T_c^0 $ and $T=T_c^0$ in a harmonic trap without the dimple and non-linear interaction. Here $T_c^0$ is the critical temperature of this boson gas. The black curve shows the density profile at $ T=0$, the blue curve shows the density profile at $T=(1/2)T_c^0 $, and the red curve shows the density profile at $T=T_c^0$. The horizontal $x$-axis is scaled by the natural length scale $a_0$ of the harmonic oscillator.}
\label{DenProfHOdT}
\end{center}
\end{figure}
The density profile $n_0(x)$ of a Bose-Einstein condensate does not depend on temperature, because it is just the absolute value square of the ground state wave function. As shown in the Figures in the previous section, all particles condense at $T=0$ according to our model as in the case of semi classical approximation. Therefore the density profile of a boson gas at $T=0$ coincides with that of a Bose-Einstein condensate. However, the density profile of a boson gas depends both on the temperature $T$ and the number of particles $N$ through the Bose-Einstein distribution and the chemical potential $\mu$. 

For the sake of completeness, we also present the density profiles of a boson gas with $N=10^4$ in a harmonic trap without the dimple potential and non-linear interaction for three different temperatures $T=0$, $T=T_c^0/2$, and $T=T_c^0$, where  $T_c^0$ denotes the critical temperature of this boson gas. Its value is calculated by Eq. \eqref{tcrit} and equal approximately $1.30$ $\mu$K for the parameters we have given previously. As is well-known, the density profile is getting widened and the peak value at the origin decreases with increasing temperature, as shown in Figure \ref{DenProfHOdT}.

\begin{figure}[h!]
\begin{center}
\includegraphics[scale=0.75]{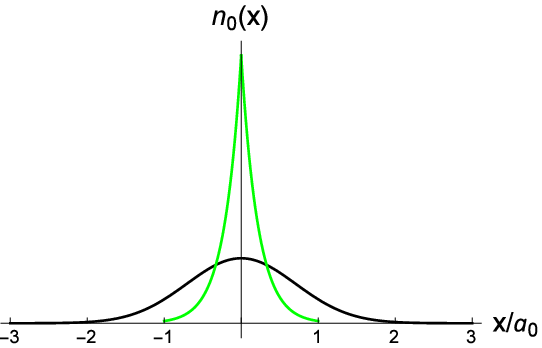}
\caption{The green curve shows the density profile of Bose-Einstein condensates in a harmonic trap with a dimple potential with $\sigma=10^6\, (\Lambda=4.6)$. The black curve shows the density profile of the Bose-Einstein condensate in a harmonic trap when the dimple potential is ignored.}
\label{DenProfs6L00}
\end{center}
\end{figure}

\begin{figure}[h!]
\begin{center}
\includegraphics[scale=0.75]{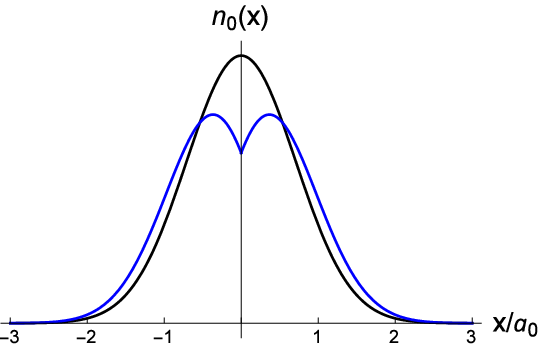}
\caption{The density profiles of Bose-Einstein condensates in a harmonic trap. The black curve shows the case when interactions are ignored $\alpha=0$ and the blue curve shows the case when the interactions around the origin are taken into account ($\alpha=10^6$). The horizontal $x$-axis is scaled by the natural length scale of harmonic oscillator $a_0=\sqrt{\hbar/(m \omega)}$.}
\label{DenProfs0L60}
\end{center}
\end{figure}

If we take into account the dimple potential, the density profile of Bose-Einstein condensate with $N=10^4$ particles for $\sigma=10^6\, (\Lambda=4.6)$ changes according to Figure \ref{DenProfs6L00}. We also show in this figure the density profile of a  Bose-Einstein condensate in a harmonic trap in the absence of the dimple potential for the sake of comparison.  We see that the presence of the dimple potential gives rise to a high concentration increase around the origin. Therefore, it is not realistic to ignore the interactions among bosons in a harmonic trap with the dimple potential.  

In Figure \ref{DenProfs0L60}, we present the density profile of condensates in a harmonic trap perturbed by the non-linear interaction (with the choice $\alpha=10^6$) in the absence of the dimple potential. If there is no dimple potential and the many-body interactions are modeled effectively by a non-linear $\delta$ potential, the peak of the density profile of the condensate at the origin is shifted symmetrically in both directions and two symmetrical peaks occur around the origin. A similar result, that shows repulsive interactions decreases the density around the origin, has also been obtained in \cite{Akram}.

\begin{figure}[h!]
\begin{center}
\includegraphics[scale=0.75]{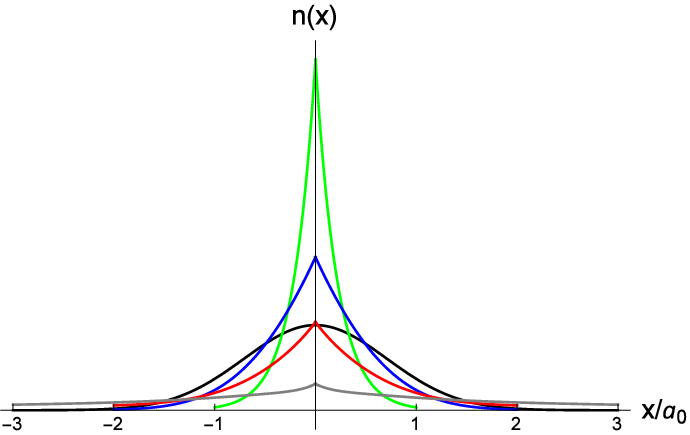}
\caption{The density profiles of boson gases at $T=0$ (indicated by the blue curve), $T=(1/2)T_c^0$ (indicated by the red curve) and $T=T_c^0$ (indicated by the gray curve) in a harmonic trap with a dimple. $T_c^0$ is the critical temperature for this potential.} \label{DenProfs6L6dT}
\end{center}
\end{figure}

However, if we consider the problem by including the dimple potential together with the non-linear delta interaction, the density profiles of boson gases in a harmonic trap perturbed by a dimple potential with $\sigma=10^6\, (\Lambda=4.6)$ and the non-linear interaction with $\alpha=10^6$ is shown in Figure \ref{DenProfs6L6dT}. In this figure, the blue, red and gray curves show the density profiles of  interacting boson gases for $T=0$, $T=T_c^0/2$ and $T=T_c^0$, respectively. We also show the density profiles of  non-interacting Bose-Einstein Condensates in a harmonic trap with and without a dimple at $T=0$ in the same figure for the sake of comparison. The green curve which has the highest peak is the density profile of a non-interacting Bose-Einstein condensate in a harmonic trap with a dimple and the black curve is the density profile of a non interacting Bose-Einstein condensate without a dimple. It can be seen from Figure \ref{DenProfs6L6dT}  that the density of the gas at the origin decreases considerably due to the interactions. Moreover, at $T=T_c^0$ the density becomes flattened except for a small peak at the origin. 

\section{Conclusions}

We have solved the bound state problem for the nonlinear Schr\"{o}dinger equation with the harmonic oscillator potential perturbed by a $\delta$ interactions, where the nonlinear term is expressed by an extra $\delta$ term proportional to $|\psi(x)|^2$. After we find the bound state energies of the system numerically, we have studied the BEC of the Bose gas by hoping to model the many body interactions among bosons by an effective concentrated nonlinear term. We have finally discussed all the thermodynamic quantities such as the critical temperature for the condensate, the condensate fraction, and the density profiles in detail and studied the effect of the nonlinear term on the condensation of the Bose gas.

\section*{Acknowledgement}

We gratefully acknowledge the anonymous referees for their suggestions which we believe they have improved our paper.

%
%


\begin{thebibliography}{99}
%
\bibitem{Demkov} Yu. N. Demkov, V. N. Ostrovskii, Zero-Range Potentials and Their Applications in Atomic Physics (Springer, Berlin, 2013).

\bibitem{Albeverio} S. Albeverio, F. Gesztesy, R. Hoegh-Krohn, H. Holden, Solvable Models in Quantum Mechanics, 2nd edn. (American Mathematical Society, Providence,
RI, 2004).

\bibitem{KronigPenney} R. de L. Kronig and W. G. Penney, Quantum mechanics of electrons in crystal lattices, \textit{Proc. Roy. Soc. A},
\textbf{130}, 499 (1931).

\bibitem{AlbeverioKurasov} S. Albeverio, P. Kurasov, Singular Perturbations of Differential Operators Solvable Schr\"{o}dinger-Type Operators
(Cambridge: Cambridge University Press, 2000).

\bibitem{Jackiw} R. Jackiw, Delta-Function Potentials in Two- and Three-Dimensional Quantum Mechanics M.A.B. Beg
Memorial Volume (Singapore: World Scientific, 1991).
%
\bibitem{Moya} P. S. Moya, M. Ramirez, M. I. Molina, Bistable transmission of plane waves across two nonlinear delta functions, \textit{Am. J. Phys.} \textbf{75}, 12, 1158 (2007).
%
\bibitem{Molina} M. I. Molina, C. A. Bustamante, The attractive nonlinear delta-function potential, \textit{Am. J. Phys.} \textbf{70}, 1, 67 (2002).
%
%
\bibitem{CMT93} D. Chen, M. I. Molina, and G. P. Tsironis, Non-adiabatic non-linear impurities in linear hosts, \textit{J. Phys.: Condens. Matter} \textbf{5}, 46, 8689–8702
(1993).
%

\bibitem{UncuErman} F. Erman, H. Uncu, Green's function formulation of multiple nonlinear Dirac $\delta$ function potential in one dimension, 
\textit{Physics Letters A}, \textbf{384} 11 126227 (2020).

\bibitem{Cacciapuoti} C. Cacciapuoti, D. Finco, D. Noja, A. Teta, The NLS equation in dimension one with spatially concentrated nonlinearities: The Pointlike limit, \textit{Lett. Math. Phys.}, \textbf{104}, 1557-1570 (2014).

\bibitem{Carlone} R. Carlone, M. Correggi, L. Tentarelli An introduction to the two-dimensional Schrodinger equation with nonlinear point interactions, \textit{Nanosystems: Physics, Chemistry, Mathematics}, \textbf{9}, 2, 187-195 (2018).

\bibitem{Cacciapuoti2} C. Cacciapuoti, D. Finco, D. Noja, A. Teta, The point-like limit for a NLS equation with concentrated nonlinearity in dimension three, \textit{Journal of Functional Analysis}, \textbf{273}, 5, 1762-1809 (2017).




\bibitem{JankeCheng} W. Janke, and B. K. Cheng, Statistical properties of a harmonic plus a delta-potential, \textit{Physics Letters A} \textbf{129}, 3, 140-144 (1988).
%
\bibitem{IoriattiRosaHipolito} L. C. Ioriatti, S. G. Rosa, and O. Hip{\'o}lito, Bose-Einstein condensation in a one-dimensional system due to an attractive-$\delta$ impurity center, \textit{American Journal of Physics} \textbf{44}, 8, 744-748 (1976).
%
\bibitem{Huncu} H. Uncu, D. Tarhan, E. Demiralp, \"{O}. E. M\"{u}stecapl{\i}o\u{g}lu, Bose-Einstein condensate in a harmonic trap decorated with Dirac $\delta$ functions, \textit{Physical Review A}, \textbf{76} 1, 013618 (2007).
%
\bibitem{Huncu2} H. Uncu, D. Tarhan, E. Demiralp, \"{O}. E. M\"{u}stecapl{\i}o\u{g}lu, Bose-Einstein condensate in a harmonic trap decorated with an eccentric dimple potential \textit{Laaser Physics }, \textbf{18}, 3, 331-334 (2008).
%
\bibitem{Grosche} C. Grosche, Path integrals for potential problems with $\delta$-function perturbation, \textit{Journal of Physics A: Mathematical and  General}, \textbf{23}, 22, 5205 (1990).
%
\bibitem{Fassari1} S. Fassari, G. Inglese, On the spectrum of the harmonic oscillator with a $\delta$ type perturbation, \textit{Helv. Physica Acta} \textbf{67}, 6, 650-659 (1994).
%
\bibitem{Fassari2} S. Fassari, G. Inglese, Spectroscopy of a three dimensional isotropic harmonic oscillator with a $\delta$ type perturbation, \textit{Helv. Physica Acta}, \textbf{69}, 2, 130 (1996).
%
\bibitem{Ersan2} E. Demiralp, Bound states of $n$-dimensional harmonic oscillator decorated with Dirac delta functions, \textit{J. Phys. A: Math. Gen.} \textbf{38}, 22, 4783 (2005).
%

\bibitem{Fassari3} S. Fassari, G. Inglese, On the spectrum of the harmonic oscillator with a $\delta$-type perturbation II, Helvetica Physica Acta, \textbf{70}, 858-865 (1997).

\bibitem{Fassari4} S. Fassari, F. Rinaldi, On the spectrum of the Schr\"{o}dinger Hamiltonian of the one-dimensional harmonic oscillator perturbed by two identical attractive point interactions, Reports on Mathematical Physics, \textbf{69},3 353-370 (2012).

\bibitem{Janev} R. K. Janev, Z. Maric, Perturbation of the spectrum of three-dimensional harmonic oscillator by a $\delta$-potential, Physics Letters A, \textbf{46}, 5 313-314, (1974).

\bibitem{Diener} R. B. Diener, B. Wu, M. G. Raizen, Q. Niu,
Quantum Tweezer for Atoms, \textit{Phys. Rev. Lett.}, \textbf{89}, 070401 (2002).

\bibitem{Stellmer} S. Stellmer, B. Pasquiou, R. Grimm, F. Schreck
Laser cooling to quantum degeneracy, \textit{Phys. Rev. Lett.}, \textbf{110}, 263003 (2013).
%
\bibitem{Parker} N. G. Parker, N. P. Proukakis, M. Leadbeater,  C. S. Adams, Soliton-sound interactions in quasi-one-dimensional Bose-Einstein condensates, \textit{Phys. Rev. Lett.}, \textbf{90}, 220401 (2003). 
%





%
\bibitem{Kurn} D. M. Stamper-Kurn, H.-J. Miesner, A. P. Chikkatur, S.Inouye, J. Stenger, W. Ketterle, Reversible formation of a Bose-Einstein condensate, \textit{Phys. Rev. Lett.}, \textbf{81}, 11, 2194
(1998).
%
%
%
\bibitem{Pinkse} P. W. H. Pinkse, A. Mosk, M. Weidemuller, M. W. Reynolds, T. W. Hijmans, and J. T. M. Walraven, Adiabatically changing the phase-space density of a trapped Bose gas, \textit{Phys. Rev. Lett.}, \textbf{78}, 6, 990 (1997).
%
\bibitem{Weber} T. Weber, J. Herbig, M. Mark, H.-C. Nagerl, and R. Grimm, Bose-Einstein condensation of cesium, \textit{Science}, \textbf{299}, 5604, 232
(2003).
%
\bibitem{Ma} Z-Y. Ma, C. J. Foot, L. Cornish, Optimized evaporative cooling using a dimple potential: An efficient route to Bose–Einstein condensation, 
\textit{J. Phys. B: At. Mol. Opt. Phys.}, 
\textbf{37}, 15, 3187-3195 (2004).

\bibitem{Comparat} D. Comparat, A. Fioretti, G. Stern, E. Dimova, B. Laburthe Tolra,
and P. Pillet, Optimized of large Bose-Einstein condensates, \textit{Phys. Rev. A}, \textbf{73}, 4, 043410 (2006).
%
\bibitem{Bason} M. G. Bason, R. Heck, M. Napolitano, O. Elíasson, R. M\"{u}ller, A. Thorsen, W-Z Zhang, J. J. Arlt, J. F. Sherson, Measurement-enhanced determination of
BEC phase transitions, \textit{J. Phys. B: At. Mol. Opt. Phys.} \textbf{51}, 175301 (2018). 
%
\bibitem{Mewes} M.-O. Mewes, M. R. Andrews, D. M. Kurn, D. S. Durfee, C. G. Townsend,  W. Ketterle, Output Coupler for Bose-Einstein Condensed Atoms, \textit{Phys. Rev. Lett.}, \textbf{78}, 582-585 (1997).
%
\bibitem{Andrews} M. R. Andres, C. G. Townsend, H.-J. Miesner, D. S. Durfee, D. M. Kurn, and W. Ketterle,
Observation of Interference Between Two Bose Condensates, 
\textit{Science}, \textbf{275}, 637-641 (1997).
%
\bibitem{Bloch} I. Bloch, T. W. H\"{a}nsch, T. Esslinger,
An Atom Laser with a cw Output Coupler, \textit{Phys. Rev. Lett.}, \textbf{275}, 3008-3011 (1999).
%
\bibitem{Chen} C-C Chen, R. G. Escudero, J. Min\'{a}\u{r}, B. Pasquiou, S. Bennetts, F. Schreck, Continuous Bose–Einstein condensation, \textit{Nature}, \textbf{606}, 683-698 (2022).
%
\bibitem{Dutta} S. Dutta, E. J. Mueller, Kinetics of Bose-Einstein condensation in a dimple potential, \textit{Phys. Rev. A}, \textbf{91}, 013601 (2015).
%
\bibitem{Akram} J. Akram, A. Pelster, Statics and dynamics of quasi one-dimensional Bose-Einstein condensate in harmonic and dimple trap, \textit{Laser Physics}, \textbf{26}, 065501 (2016).
%
\bibitem{Karabulut} E. \"{O}. Karabulut, Effect of a dimple potential on the ground-state properties of a quasi-one-dimensional Bose–Einstein condensate with two-and three-body interactions, \textit{Physica B}, \textbf{462}, 104-111 (2015).
%
\bibitem{Sakhel} R. Sakhel, A. R. Sakhel, H. B. Ghassib, and Antun Balaz, Conditions for order and chaos in the dynamics of a trapped Bose-Einstein condensate in coordinate and energy space, \textit{Eur. Phys. J. D}, \textbf{66} (2016).

\bibitem{lebedev} N. N. Lebedev and R. A. Silverman, D. B. Livhtenberg, \textit{Special Functions and Their Applications}, Dover Publications, New York, (1965).

%
\bibitem{GradRyznik} I. S. Gradshteyn and I. M. Ryzhik, \textit{Table of integrals, series, and products}, Academic press, 2014.
%
\bibitem{hau} L. V. Hau, S.E. Harris, Z. Dutton, and C.H. Behroozi, Light speed reduction to 17 metres per second in an ultracold atomic gas, \textit{Nature}, \textbf{397}, 594-598 (1999).
%
\bibitem{Garret} M. C. Garrett, A. R. Eikbert, D. van Ooijen, C. J. Vale, K. Weegink, S. K. Schnelle, O. Vainio, N. R. Heckenberg, H. Rubinsztein-Dunlop, M. J. Davis,
Growth dynamics of a Bose-Einstein condensate in a dimple trap without cooling, \textit{Physical Review A}, \textbf{83}, 1, 013630 (2011).
%

\bibitem{Görlitz} A. Görlitz, J. M. Vogels, A. E. Leanhardt, C. Raman, T. L. Gustavson, J. R. Abo-Shaeer, A. P. Chikkatur, S. Gupta, S. Inouye, T. Rosenband, and W. Ketterle, Realilzation of Bose-Einstein condensates in lower dimensions,  \textit{Phys. Rev. Lett.}, \textbf{87}, 13, 130402  (2001).

\bibitem{Erman1} K. G. Akba\c{s}, F. Erman, O. T. Turgut, On Schr\"{o}dinger operators modified by $\delta$ interactions, \textit{Annals of Physics}, \textbf{458}, 2, 169468 (2023).  
%
%
\bibitem{pethick} C. J. Pethick and H. Smith, \textit{Bose Einstein Condensation in Dilute Gases}, Second edition, Cambridge University press, New York, (2008). 
%


%
%
%
%

\end{thebibliography}
\end{document}